\documentclass[openacc]{rstransa}

\jname{rsta}
\Journal{Phil. Trans. R. Soc}
\begin{document}

\title{Time-Energy Uncertainty Principle for Irreversible Heat Engines}

\author{
Rudolf Hanel$^{1}$ and Petr Jizba$^{2}$}

\address{$^{1}$Section for Science of Complex Systems, Medical University of Vienna, Spitalgasse 23, 1090 Vienna, Austria\\
$^{2}$FNSPE, Czech Technical University in Prague, B\v{r}ehov\'{a} 7, 115 19, Praha, Czech Republic}

\subject{xxxxx, xxxxx, xxxx}


\corres{Petr Jizba\\
\email{p.jizba@fjfi.cvut.cz}}

\begin{abstract}
Even though irreversibility is one of the major hallmarks of any real life process, an actual under- standing of irreversible processes remains still mostly semi-empirical.
In this paper we formulate a thermo- dynamic uncertainty principle for irreversible heat engines operating with an ideal gas as a working medium.
In particular, we show that the time needed to run through such an irreversible cycle multiplied by the irreversible work lost in the cycle, is bounded from below by an irreducible and process-dependent constant that has the dimension of an action. The constant in question depends on a typical scale of the process and becomes comparable to Planck's constant at the length scale of the order Bohr-radius, i.e., the scale that corresponds to the smallest distance on which the ideal gas paradigm realistically applies.
\end{abstract}


\begin{fmtext}
\section{Introduction}
Our daily experience shows us that most processes around us happen irreversibly.
Sugar, once dissolved in our morning coffee, does not spontaneously reconstitute itself, and coal, once combusted
in the open air, does not spontaneously reassemble into barbeque charcoal.
Though there are various fully reversible processes at the atomic and subatomic levels,
there are none at the macro scale. No large-scale process is
perfectly reversible,
since
at least small bits of energy get lost from a system whenever it transforms energy.
Irreversible, dissipative processes are governing our lives so ubiquitously that it may
come as a surprise how little we actually understand theoretically about them. This remains
true, despite the considerable amount of scientific work dedicated to non-equilibrium thermodynamics~\cite{netd}.

\end{fmtext}
\maketitle
\noindent

Non-equilibrium thermodynamics deals with such phenomena as coupled transport processes~\cite{Onsager}
or finite-speed heat engines working between heat baths with finite heat transfer coefficients~\cite{CurzonAhlborn}, just to name a few. Many theoretical approaches, such as superstatistics~\cite{BeckCohen,HTMGM1}, classical irreversible thermodynamics~\cite{Onsager,Meixner,Jou1988}, stochastic thermodynamics~\cite{Seifert2008} or the thermodynamics of small systems~\cite{Rossnagel2014},
typically rely on the \textit{local equilibrium} or \textit{stationarity}
assumptions~\cite{Onsager,Prigogine,HafskoldKjelstrup}.
On the other hand, for systems working between an energy potential, e.g. a hot and a cold reservoir, which drives a continuous energy current through the system, the local equilibrium assumption is typically violated, and often one has to resort to numerical simulations as the primary (and often the only) diagnostic tool~\cite{Chandrasekhar,CrossHohenberg}, despite recent advances in the theory of driven systems \cite{Van,HTmixedEntropy}.

Much of what we seem to understand about irreversible processes comes from semi-empirical considerations
rather than from first principle derivations. In fact, we find ourselves in a rather awkward position, since
the theory we understand best, namely reversible thermodynamics, cannot easily
be adapted to the description of irreversible processes without sacrificing the equilibrium
concepts, i.e. the very concepts on which reversible thermodynamics fundamentally hinges.
One might even go as far as to say that the undeniable success of quasi-equilibrium theory of
reversible processes has been instrumental in obscuring intrinsic mechanisms responsible for
thermodynamic irreversibility.
Here we attempt to offer an explicit insight into how {\em mechanical} irreversibility of thermodynamic processes
can be understood for an \textit{ideal gas} system (i.e. a system consisting of non-interacting point-like particles)
and elastic collisions with the confining vessel boundaries, based
on a more than half a century old observation, that momentum transfer between molecules and moving piston,~\cite{LST_StatTherm}, produces irreversible work contributions~\cite{Bauman}.
In particular, our aim will be to first derive the path-dependent work equation for this
important contribution to the irreversibility; a contribution
that is entirely due to the motion of the piston rather than being the consequence of non-equilibrium
processes happening inside the bulk of the gas. Secondly, we shall use this result to derive a ``thermal uncertainty relation'' connecting minimal irreversible energy requirements of a process with the time it takes to perform the process,
demonstrating that running processes faster comes with a penalty --- an increased minimal irreversible energy cost.
%
For that reason our discussion will focus only on mechanical-interface-induced irreversibility, while, at the same time, we will neglect friction or finite heat transfer between heat bath(s) and working medium. This will, on one hand simplify our technical discussion but at the same time it will provide us sufficiently versatile playground that will allow us to address some of the salient features that are key for understanding irreversible behaviour of generic heat engines.

\begin{figure}[h]
\centering\includegraphics[width=2.5in]{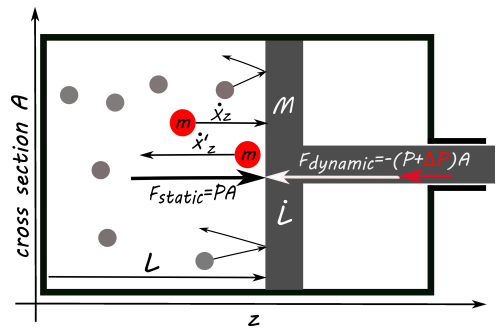}
\caption{Irreversible processes are also {\em out of equilibrium} processes. The irreversibility of
thermodynamic cycles caused at the mechanical interface, in this example the piston controlling the volume of a ideal-gas-filled cylinder, is mainly caused by the extra pressure $\Delta P$ that one has to apply to move the piston at non-zero speed. During compression one requires $\Delta P>0$, during expansion  $\Delta P<0$, which leads to a slight deviation of the force the gas exerts on the static cylinder walls or the static piston in comparison to a piston that is moving. $F_{\rm dynamic}$ therefore drives the system through a cycle in finite time, but also breaks homogeneity and isotropy of the gas-particle and gas-particle velocity distribution in the cylinder, however slightly; Even if we assume
that in the bulk of the gas the particles and their velocities remain homogeneously and isotropically distributed,
the velocities of the particles rebounding from the piston, which carry additional momenta added by the moving piston in the collisions, will break the isotropy of the particle velocity distribution.}
\label{fig1}
\end{figure}

Because of its simplicity, the \textit{ideal gas} represents a quintessential system of reversible thermodynamics.
It may thus come as a surprise that within the ideal gas paradigm one can quite easily attack issues
related to irreversible thermodynamics. In fact, the only thing that is required in this context is to understand
the ``mechanical interface'', namely the dynamics of the piston  that controls the volume of the ideal gas confined in
a cylindrical vessel as it moves with some phenomenologically relevant non-zero speed. We consider a
cylindrical container merely for a technical convenience and results obtained are by no means restricted to this
particular shape.
%
For modelling the mechanical interface
one has to consider the statistics of elastic collisions between gas particles and piston, leading to relations between macro observables. We should stress that apart from the usual macro variables such as
$N$ (particle number), $P$ (pressure), $V$ (volume),
and $T$ (temperature of the working medium), we also have to consider other
macro variables such as the rate of the change of the volume $\dot V$.
Here, the volume, $V=A L$, is the product of $A$, the cross-section area of the cylinder confining the gas,
with $L$, the axial cylinder dimension, see Fig.~\ref{fig1}.
Note that by considering $\dot V$ as an additional thermodynamic state
variable brings about an explicit violation of the concept of local equilibrium. The latter is also an essential point of
%
Extended Irreversible Thermodynamics~\cite{netd,Jou1988} and Rational Extended Thermodynamics~\cite{Ruggeri}.

The aforementioned will suffice to show that irreversibility of a mechanical work has intimate connections with processes happening at finite speed. The surprising result of this paper is that the amount of mechanical work we loose irreversibly per single degree of freedom, i.e. $\Delta w_{\rm irrev}$, within the time period $\tau$ satisfies the following time-energy
``\textit{uncertainty relation}''
\begin{equation}
\Delta w_{\rm irrev}\tau \ \geq  h_{\rm process}\,.
\label{uncert1}
\end{equation}
The total work lost irreversibly would then be $\Delta W_{\rm irrev}=Nf \Delta w_{\rm irrev}$, where
$N$ is the number of particles
and $f$ the number of particle degrees of freedom, and $h_{\rm process}>0$ is a process-dependent
constant that has the dimension of an action.
The value of the constant $h_{\rm process}$ depends on the physical scale of the process.
As will be shown, $h_{\rm process}$ is essentially bounded from below by
Planck's constant,
which makes a surprising parallel to the Heisenberg time-energy uncertainty relation, despite the
fact that both uncertainty relations have very different conceptual origins. It should perhaps be noted that in our reasoning we do not use any quantum mechanical but purely classical mechanical considerations.

Aforementioned uncertainty relation allows one to regain an intuition that seems to have vanished from classical equilibrium thermodynamics. Loosely speaking Equ. (\ref{uncert1}) can be interpreted as follows;
to run a process faster we inevitably loose more energy irreversibly, meaning that we cannot recover this energy by running the process in reversed direction. Consequently, faster turning heat
engines become less efficient. So, to perform work faster makes it inevitably less efficient.
One particularly important implication for heat engines that follows from Equ. (\ref{uncert1}),
is the existence of an upper bound for its rate of change, the so called \textit{idle speed}, which is reached when the process runs so fast that its efficiency becomes zero. \vspace*{-7pt}

The paper is organized as follows. In the next section (2) we set up the ideal gas model that will be instrumental in discussing irreversible work.
In particular we will discuss the statistics of elastic collisions between particles and piston and the ensuing average particle velocity sampled by the piston.
As a next step we derive a path-dependent work
functional, then
determine the optimal path through a variational
principle,
and finally derive the equation relating the macro state variables. After this preliminary work we analyse in section (3) the isothermal processes and set up the associated uncertainty relation and action constant. Similarly, in section (4) we study irreversible adiabatic processes and find the associated work-time uncertainty relation. Carnot-like irreversible heat engines are finally analysed in section (5),
where we combine the results of the previous two chapters to compute the uncertainty relation for irreversible
cyclic processes. We conclude with discussing the efficiency and power output of such processes and their prospective applications.
Some finer technical steps are relegated to Supplemental Material (SM).

%
%
%
%
%
%
%
%
\section{The ideal gas model of irreversible work}

Here we will briefly discuss the ideal gas model employed in this paper. Let us consider a cylinder with a cross section area $A$ oriented so that its axis points into $z$-direction.
The left end of the cylinder is closed and the right end is controlled by a piston, which moves in z-direction. The
piston position $L$ on the $z$-axis corresponds to its distance to the closed end of the cylinder
so that the enclosed gas volume equals $V=AL$. We consider $L$ and its time derivative $\dot L$ as macro-variables, corresponding to a local time average of the actual piston position and velocity respectively. At a microscopic scale these will slightly fluctuate.
Directly after a collision with a molecule the piston will have a velocity
$\underline{u}(t)$ while directly before the collision the velocity was $\overline{u}(t)$.
In the timespan $\Delta t$ between collisions the piston gets accelerated by an external force
$F_z(t)=Ma(t)$, where $M$ is the piston mass and $a(t)$ is the acceleration of the piston in the
microscopic time interval $[t,t+\Delta t]$, i.e.:
\begin{equation}
\overline{u}(t+\Delta t) \ = \ \underline{u}(t)\ + \ a(t)\Delta t\, .
\end{equation}
We assume that $a(t)$ is approximately constant between two successive collisions
and for simplicity's sake we also assume that $\Delta t$ is so short that only single particle collisions
take place within this time window. We also define the macro-variable $\dot L$ as
\begin{equation}
\dot L(t) \ = \ \frac{1}{2}[\underline{u}(t)\ + \ \overline{u}(t+\Delta t)]\,.
\end{equation}

In fact, the piston is a thermodynamic system of its own, but for our purposes we will consider it to be one rigid block with constant internal energy playing no role in any of our following considerations, thereby reducing it to one degree
of freedom (according to our convention in the $z$-direction) that can be controlled externally. The latter further implies that the $x$ and $y$ particle velocity components do not explicitly play any role as the piston can recoil in collision events only in the $z$-direction.
The inherent jitter of the piston position and its velocity around their respective macroscopic values,
reflecting the piston temperature will be neglected in the following.
This simplification will allow us to demonstrate the mathematical reasons of mechanical irreversibility
without considering details that would unnecessarily complicate our discussion.
For the same reason we assume that the piston glides without friction within the cylinder
and that heat transfer between possible heat-baths and the working medium
(ideal gas) in the piston happens quasi-instantaneously, i.e. the gas temperature can still be set to the heat bath temperature.
This would not be true if finite heat transfer coefficients would cause a lag between the working medium
and heat-bath temperature in dynamical situations~\cite{CurzonAhlborn,MatolcsiT}. Similarly, our ideal gas model ignores well know technical sources of irreversibility, such as convective losses in the working medium. We also ignore that external gas molecules act on the piston, i.e. we assume that the cylinder and
the piston are placed in a vacuum. With those simplifying assumptions we not only reduce the system to its basic components, i.e. gas molecules and piston, which assume a statistical description in terms of classical mechanics, but
we also eliminate most of non-mechanical effects that are known to contribute to irreversibility.
Still, what remains is a non-negligible source of work that has to be spent irreversibly in any act of compression or expansion. At the so called \textit{idle speed} the efficiency of  a heat  engine becomes zero
and all work produced by the engine is spent on running it.
We relegate the discussion of this point to section (\ref{sec:carnot}).
%

\begin{figure}[h]
\centering\includegraphics[width=2.7in]{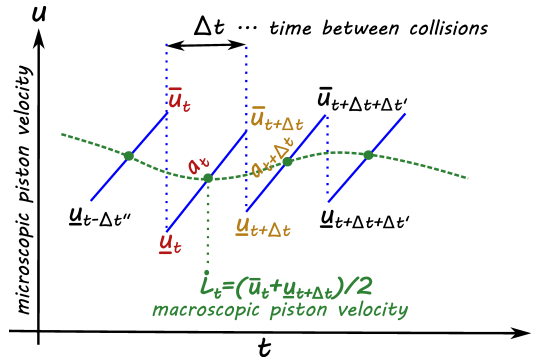}
\caption{The time evolution of the piston velocity is sketched.
	On a very fine time resolution $\Delta t$ one may assume that the piston experiences only single particle collisions,
	and as a consequence, once a gas particle collides with the piston, the next particle will on average
	collide with the piston  a time $\Delta t$ later.
	The piston therefore undergoes a microscopic dynamics that re-accelerates the piston from a velocity
	$\underline{u}_t$ the piston assumes immediately after a collision, to the velocity $\overline{u}_{t+\Delta t}$
	the piston assumes directly before the next collision, corresponding to a local average velocity $\dot L_t$.
	Note that for better readability we identify $L_t\equiv L(t)$, $\underline{u}_t\equiv\underline{u}(t)$, and
	 $\overline{u}_t\equiv \overline{u}(t)$.
	}
\label{fig2}
\end{figure}


By employing momentum and energy conservation we can now compute the effect of the elastic collision on the microscopic piston velocities $u(t)$ and its associated values $\underline{u}(t)$ and $\overline{u}(t)$.
One can easily check that
\begin{equation}
\underline{u}(t) \ = \ \frac{2m}{M+m}v_z \ + \ \frac{M-m}{M+m}\overline{u}(t)\, ,
\end{equation}
where $m$ is the gas particle mass and $v_z$ is the gas particle velocity in $z$-direction.
%
The rebounding velocity is given by
\begin{equation}
v'_z \ = \ \frac{2M}{M+m}\overline{u}(t)\ - \ \frac{M-m}{M+m}v_z\,.
\label{reboundvelo}
\end{equation}
By employing the fact that we control the macroscopic force $F_{\rm instant}=Ma(t)$ smoothly on macroscopic time scales,
so that for two subsequent collisions it is (on average) true that
$a(t+\Delta t)\Delta'\sim a(t)\Delta t$,
where $\Delta t$ is the time elapsing between first and second collision  at time $t$ and $t+\Delta t$
and $\Delta'$ is a the time elapsing between the second collision and the third one at time $t+\Delta t+\Delta t'$
and further, that $\ddot L(t)=(\dot L(t+\Delta t)-\dot L(t))/\Delta t$, we obtain
\begin{eqnarray}
\frac{2m}{M+m}\left[v_z-\dot L(t)\right] \ = \ \left[\ddot L(t)-\frac{M}{M+m}a(t)\right]\Delta t\,,
\label{collequ}
\end{eqnarray}
with $\Delta t$ being the time span elapsing between collisions
(compare SM~\cite{SM}).
Equivalently, we can write $2m[v_z-\dot L(t)]^2 =  [(M+m)\ddot L(t)-Ma(t)]\Delta z$,
where $\Delta z = [v_z-\dot L(t)]\Delta t$ represents the
inter-particle distance in the $z$ direction,
for particles with the corresponding velocity component $v_z$.
The origin of history dependence in our thermal system can be at this point retraced
to the left hand side of Eq.~(\ref{collequ}), which clearly breaks the time-reversal symmetry of
the moving piston.
In order to proceed further with our statistical reasoning
we need to estimate the piston average particle velocity $\langle v_z \rangle_{\rm piston}$,
which is taken with respect to the particle velocity distribution sampled by the piston,
and the characteristic inter-collision time $\overline{\Delta t}$.
We should stress that the latter is not of Maxwell--Boltzmann type
(compare SM~\cite{SM}).

\subsection{The statistics of collisions}

For piston velocities $\dot L$ that are much smaller than the
typical gas particle velocities, we can still employ
the
equipartition identity
\begin{equation}
\frac{1}{2}m \bar v_z^2 \ = \ \frac{1}{2} k_B T \ = \ \frac{1}{f}\frac{U}{N}\, ,
\end{equation}
as the average kinetic energy per degree of freedom,
where $f$ is the number of particle's internal degrees of freedom (e.g. $f=3$ for a mono-atomic gas),
$\beta=1/k_BT$ is the inverse temperature, $k_B$ is the Boltzmann constant,
and $T$ the temperature of the working gas. Similarly,
$U$ is the {\em internal energy} of the system and $\bar v_z$ is the \textit{equipartition velocity} in $z$-direction,
i.e.
\begin{equation}
\bar v_z \ = \ \sqrt{\frac{k_BT}{m}}\quad {\rm or}\quad  \bar p_z \ = \ \sqrt{mk_BT}\, ,
\end{equation}
where $\bar p_z=m\bar v_z$ is the ensuing particle's \textit{equipartition momentum}.

To a first approximation, one could assume
that $\langle v_z\rangle_{\rm piston}=\bar v_z$. In this case one would conclude that
the average particle distance of particles that are  heading towards the piston is given by $\overline{\Delta z}=2L(t)/N$.
Similarly, the average time elapsing between two collisions  needs to
be $\overline{\Delta t}=\frac{1}{N}\frac{2L(t)}{\bar v_z-\dot L(t)}$.
Corrections beyond this approximation are provided
in the form of a perturbation expansion that is phrased in terms of
dimensionless parameter $\dot L/\bar v_z$,
respecting thus the relative velocity distribution sampled by the moving piston.

\subsection{Work and the equation of states}

The total work that is needed to move the piston by a distance $dL(t)$
(and also compressing/expanding the gas) is given by
$dW(t)=Ma(t)dL(t)$. The work $W(t)$ can be computed by inserting the estimates for
$\overline{\Delta z}$ and $\langle v_z\rangle_{\rm piston}$ into Eq.~(\ref{collequ}). Consequently,
to the first order in $\dot L/\bar v_z$, we have (cf. SM~\cite{SM})
\begin{equation}
\dot W(t) \ = \ M\ddot L(t)\dot L(t ) \ + \ \dot W_{\rm gas}(t)\,,
\label{workfunctionalA}
\end{equation}
with
\begin{equation}
\dot W_{\rm gas}(t) \ = \ - Nm \frac{\dot L(t)}{L(t)} \bar v_z^2\left[1 \ - \ 2q\frac{\dot L(t)}{\bar v_z} \right]\,,
\label{workfunctionalC}
\end{equation}
where $q=\sqrt{2/\pi}$ (see SM). To move the piston, the total force $Ma(t)$ needs to be applied.
The term $M\ddot L$ in Eq.~(\ref{workfunctionalA}) describes the
force corresponding to the reversible macroscopic acceleration
of the piston mass.
The kinetic energy $M\dot L^2/2$ stored in the piston velocity can, however, in principle, be reversibly
retrieved (i.e. $\oint dL \ddot L=0$) for all closed paths $L$.
On the other hand the
force component
$F_z=N  m  \bar v_z^2\left(1 - 2q \dot L/\bar v_z \right)/L$
describes the gas pressure, $P_z=F_z/A>0$,  against the moving piston, which has reversible \textit{and}
irreversible components.
It should be stressed that the gas pressure on the static cylinder walls is given
by $P=F/A$ with
$F=N m \bar v_z^2/L$,
which is the fully reversible expression, as expected.
One gets two sets of state equations. One for the moving piston,
\begin{equation}
P_zV \ = \  N m  \bar v_z^2\left[1 \ - \ 2q\frac{\dot L(t)}{\bar v_z} \right],
\label{equstateA}
\end{equation}
and one for the static cylinder walls, i.e. the static piston ($\dot L=0$)
\begin{equation}
PV \ = \  N m \ \bar v_z^2\, .
\label{equstateB0}
\end{equation}
As a direct consequence of the last two equations it is easy to identify the pressure difference
$\Delta P=P_z-P$ as
\begin{equation}
\frac{\Delta P}{P}\ = \  - \, 2q\frac{\dot L(t)}{\bar v_z} \,,
\label{pressuredifference}
\end{equation}
This result essentially coincides with the classic result of Bauman and Cockerham~\cite{Bauman} even though their methodology is substantially differing from ours.
The static equation of states, Eq.~(\ref{equstateB0}), reduces to the equilibrium equation of states of the ideal gas
\begin{equation}
PV \ = \  Nk_BT\,.
\label{equstateB}
\end{equation}

\subsection{Variation of work $W_{\rm gas}(t)$ over possible histories}

Let us now ask the question, how much work do we need to invest into compressing
the gas from $V_1$ to $V_2<V_1$? Since work given in Eq. (\ref{workfunctionalC})
is history-dependent the previous question cannot be answered without
specifying the history, that is the path $t\to L(t)$ of the piston.
Our goal is to specify histories that minimize irreversible energy losses.
We do this in two steps. We first discuss isothermal processes and after this we turn to the adiabatic case.

Our strategy is based on a variational approach.
In particular we search for such histories that extremise the irreversible work
loss $\delta W_{\rm gas}=0$ for
\begin{equation}
W_{\rm gas} \ = \ -Nm \bar v_z^2 \int_{t_0=0}^{t_1=\tau}dt \frac{\dot L(t)}{L(t)}\left( 1-2q\frac{\dot L(t)}{\bar v_z} \right)\,,
\label{workfunctionalCx}
\end{equation}
which is merely the integral of Eq.~(\ref{workfunctionalC}).
Note, that for isothermal work $\bar v_z$ is a history independent constant.
The variational principle gives us (in our first order approximation) the differential equation
%
\begin{equation}
\frac{\dot L}{L}\ = \ 2 \ \! \frac{\ddot L}{\dot L}\, ,
\label{isovarprincA}
\end{equation}
with the solution
\begin{equation}
L(t) \ =  \left[ \sqrt{L_1}+\frac{t}{\tau}\left(\sqrt{L_2}-\sqrt{L_1}\right)\right]^2 ,
\label{isovarprincBsol}
\end{equation}
where $L_1$ is the initial piston position at $t=0$ and $L_2$ the end position at $t=\tau$.

Interestingly, also for the adiabatic work, the variational principle yields the same result for $L(t)$.
This is a direct consequence of that fact that in the adiabatic processes $dW(t)=dU(t)$
and $\bar v_z(t)=\sqrt{2U(t)/fNm}$. The variation of the ensuing work $\delta W_{\rm gas}$
involves terms
\begin{equation}
\delta \bar v_z(t) \ = \   \frac{\delta W_{\rm gas}(t)}{fNm\bar v_z(t)}\,.
\label{adiavarprinc}
\end{equation}
Since we require $\delta W_{\rm gas}(t)=0$ for all $t$ it then also implies that $\delta \bar v_z(t) =0$.
As a consequence, the adiabatic variational principle yields the same solution as the isothermal case,
as already stated above.

In passing, it should be noted that there exists no smooth (non-singular) solution for the above variational principle that satisfies the constraints $\dot L(t_0) = \dot L(t_1) = 0$.
This in turn implies that it is impossible to construct a machine that actually attains minimal irreversible energy losses and is still compatible with those boundary conditions.

\section{Isothermal irreversible work}

To obtain the isothermal work we have to integrate Eq.~(\ref{workfunctionalC}) for constant
$\bar v_z$ and for a path $L(t)$ given in Eq.~(\ref{isovarprincBsol}) that begins in $L_1$ at time $t_1=0$ and
ends in $L_2$ at $t_2=\tau$.
In doing so we obtain
\begin{equation}
\Delta W_{\rm gas}|_1^2(\tau) \ = \ \Delta W_{\rm rev}|_1^2(\tau) \ + \ \Delta W_{\rm irrev}|_1^2(\tau)\,,
\label{isowork2a}
\end{equation}
with
\begin{equation}
\Delta w_{\rm rev}|_1^2(\tau) \ = \ -\frac{ m }{f} \ \! \bar v_z^2 \ \!\left[\log(L_2) \ - \ \log(L_1)\right]\,,
\label{isowork2b}
\end{equation}
and
\begin{equation}
\Delta w_{\rm irrev}|_1^2(\tau) \ = \ \frac{8q m }{\tau f} \ \! \bar v_z \left[\sqrt{L_2} \ - \ \sqrt{L_1}\right]^2\,,
\label{isowork2c}
\end{equation}
where we have denoted with  $w=W/(fN)$ the work per degree of freedom.
Eq.~(\ref{isowork2c}) can be conveniently rewritten into the form
\begin{equation}
h_z \ \equiv \  \bar p_z D_z(L_1,L_2) \ = \ 4 \tau \Delta w_{\rm irrev}|_1^2(\tau)\, ,
\label{isouncert}
\end{equation}
where we have defined $h_z$ to be the \textit{isothermal action} and
\begin{equation}
D_z(L_1,L_2)\ = \ \gamma_z g(L_1,L_2)\, ,
\end{equation}
is the \textit{characteristic length scale} of the process,
with a \textit{coupling constant}
\begin{equation}
\gamma_z \ = \  64\frac{q}{f} \quad {\rm and }\quad g(L_1,L_2) \  = \  \frac{1}{2} \left[ \sqrt{L_1}-\sqrt{L_2} \right]^2\ .
\end{equation}
The latter can also be written as
\begin{equation}
g(L_1,L_2) \ = \  \frac{1}{2} \left[L_1 \ + \ L_2 \right] \ - \ \sqrt{L_1L_2}\, ,
\end{equation}
which is nothing but the difference between the arithmetic and the geometric mean of $L_1$ and $L_2$.
In passing we note that the notion \textit{characteristic length scale} for $D(L_1,L_2)$ is motivated by the
close analogy of Eq. (\ref{isouncert}) with Heisenberg position-momentum uncertainty relation.

One notes that $\Delta W_{\rm rev}|_1^2(\tau)=-\Delta W_{\rm rev}|_2^1(\tau)$
(as expected from reversible work) but
$\Delta W_{\rm irrev}|_1^2(\tau)=\Delta W_{\rm irrev}|_2^1(\tau)>0$ for all $\tau<\infty$ and $|L_2-L_1|>0$.
Let us now perform a isothermal cycle
$L_1\ \stackrel{\tau/2}{\rightarrow}\ L_2\ \stackrel{\tau/2}{\rightarrow}\ L_1$ with a cycle period $\tau$.
The corresponding irreversible work over such a cycle is
\begin{equation}
\Delta W_{\rm isothermal}(\tau) \ \equiv \ \Delta W_{\rm gas}|_1^2(\tau/2) \ + \ \Delta W_{\rm gas}|_2^1(\tau/2)\,,
\label{isowork2d}
\end{equation}
with
\begin{equation}
h_z \ = \  \bar p_z D_z(L_1,L_2) \ = \ \tau \Delta w_{\rm isothermal}(\tau)\,.
\label{isowork2dx}
\end{equation}
%
As a consequence, we find that for a general isothermal process, i.e. an isothermal process with
a generic (not necessarily variational) path $L(t)$ through the cycle, it follows that
\begin{equation}
\Delta W_{\rm isothermal}(\tau)\ \geq \  fN h_z/\tau \ > \ 0\,,
\label{isoprinciple}
\end{equation}
with equality if and only if $L(t)$ is given by (\ref{isovarprincBsol}).
Thus, the irreversible work per gas molecule that is lost in the isothermal cycle
times the time it takes to run through the cycle is always larger (or at best equal) to a
lower positive bound $h_z$ with the dimension of an action. Note that this constant only depends on the temperature $T$
and the particle mass $m$, via $\bar p_z$, and the boundary points of the cycle $L_1$ and $L_2$ via the
characteristic length scale $D(L_1,L_2)$.

Result (\ref{isoprinciple}) implies that the irreversible power consumption of an isothermal
cycle becomes
\begin{equation}
{\cal P}_{\rm isothermal} \ \geq \ fN\frac{h_z}{\tau^2} \ > \ 0\,,
\label{isopowcyc}
\end{equation}
which diverges like $\tau_{\rm cycle}^{-2}$ as $\tau_{\rm cycle}\to 0$.
Note that the positive value of ${\cal P}_{\rm isothermal}$ means that dissipative work has been performed on the system and that this energy is irreversibly lost in form of heat absorbed by the heat bath.

\section{Adiabatic irreversible work}

Unlike the isothermal processes the adiabatic process does not allow for any
heat flow between the heat bath and the system, but the work, $dW_{\rm gas} = dU$,
which corresponds to the energy transfer between piston an gas molecules via elastic collisions,
simply adds to the internal energy $U$ of the system.
If we now use that $U(t)=fN\frac12m\bar v_z(t)^2$
together with Eq.~(\ref{workfunctionalC}), we obtain the differential equation
\begin{equation}
\frac{d}{dt} \bar v_z(t) \ = \ -\frac{1}{f}\frac{\dot L}{L}\bar v_z(t) \ + \     \frac{2q}{f}\frac{\dot L^2}{L}\,.
\label{adiabatic1}
\end{equation}
This equation can be easily integrated for the path $L(t)$ given by Eq.~(\ref{isovarprincBsol}), for the boundary
piston positions $L_1$ (at time $t_1=0$) and $L_2$ (at time $t_2=\tau$). For this path one gets
\begin{equation}
\bar p_z(T_2)  L_2^{\alpha} \ = \  \bar p_z(T_1) L_1^{\alpha} \ + \ \frac{m}{4\tau}D_z^*(L_1,L_2)\, ,
\label{adiabatic2}
\end{equation}
%
where $\alpha= 1/f $, and
\begin{equation}
D_z^*(L_1,L_2)\ = \ \gamma_z g^*(L_1,L_2)\, ,
\label{adiacharscale}
\end{equation}
is the \textit{characteristic length scale} of the adiabatic process.
Moreover, $g^*(L_1,L_2)=g(L_1,L_2)\lambda(L_1,L_2)$, with
\begin{equation}
\lambda(L_1,L_2) \ = \ \frac{1}{1+2\alpha}\frac{ L_1^{\alpha+\frac{1}{2}}-L_2^{\alpha+\frac{1}{2}} }{ \sqrt{L_1}-\sqrt{L_2} } \, .
\label{adiametric}
\end{equation}
Moreover, we define $\lambda(L_1|L_2)=\lambda(L_1,L_2)L_2^{-\alpha}$  and
$D_z^*(L_1|L_2)=D_z^*(L_1,L_2)L_2^{-\alpha}$.
By employing the notation $u=U/fN$ for the internal energy per degree of freedom we can write
the irreversible change of the internal energy $\Delta u_{\rm adiabatic}(\tau)=\Delta u|_1^2(\tau/2)+\Delta u|_2^1(\tau/2)$ for the adiabatic cycle $L_1\stackrel{\tau/2}{\rightarrow} L_2 \stackrel{\tau/2}{\rightarrow} L_1$ in the following thermodynamic uncertainty relation like form,
\begin{equation}
h_z^* \ \equiv \ \bar p_z D_z^*(L_2|L_1) \ = \ \tau \left(\Delta u_{\rm adiabatic}(\tau) \ - \ \varepsilon(L_2|L_1;\tau) \right),
\label{adiauncertainty}
\end{equation}
where we call $\varepsilon$ the irreversible adiabatic offset energy
\begin{equation}
\varepsilon(L_2|L_1;\tau) \ = \ \frac{1}{2}m\left[\frac{D^*(L_2|L_1)}{\tau}\right]^2.
\label{adiaoffset}
\end{equation}
Note that adiabatic irreversible work not only has a term that
diverges for small $\tau$ as $1/\tau$ but another one that diverges as $1/\tau^2$.
The reduced work $u|_1^2$ for the adiabatic process, $(L_1,T_1,t_1)\to (L_2,T_2,t_2)$ in the time $\tau=t_2-t_1$,
is then given by
\begin{equation}
\Delta u|_1^2(\tau)\ = \ \Delta u_{\rm rev}|_1^2(\tau) \ + \ \Delta u_{\rm irrev}|_1^2(\tau)\,,
\label{adiawork12}
\end{equation}
with
\begin{equation}
\Delta u_{\rm rev}|_1^2(\tau) \ = \ \frac{1}{2}k_B T_1\left[ \left( \frac{L_1}{L_2} \right)^{2\alpha}-1 \right]\,,
\label{adiawork12rev}
\end{equation}
not explicitly depending on the time $\tau$ (which is the hallmark of reversibility), and
\begin{equation}
\Delta u_{\rm irrev}|_1^2(\tau) \ = \ \frac{1}{4\tau} p_z D^*(L_2|L_1) \ + \ \frac{1}{16}\varepsilon(L_2|L_1;\tau)\,.
\label{adiawork12irrev}
\end{equation}\vspace*{-7pt}


Let us note that the equilibrium analogue of Eq.~(\ref{adiawork12rev}) looks identical in form with the quasi-static reversible adiabatic work provided that $2\alpha=\kappa -1 $, where $\kappa$ is the Poisson constant. In equilibrium this means that $\kappa=C_p/C_v$ and hence $\kappa-1=2/f$. The slow piston assumption does not alter this relation ($\alpha=1/f$).
However, we want to point out that corrections to this constant may become necessary, if one considers extreme changes in the force $M a(t)$ applied to the piston, strong enough to change $a(t)$ on microscopic time scales,
so that $a(t)\Delta t \sim a(t+\Delta t)\Delta t'$ is no longer true on overage.

\begin{figure}[h]
\centering\includegraphics[width=5in]{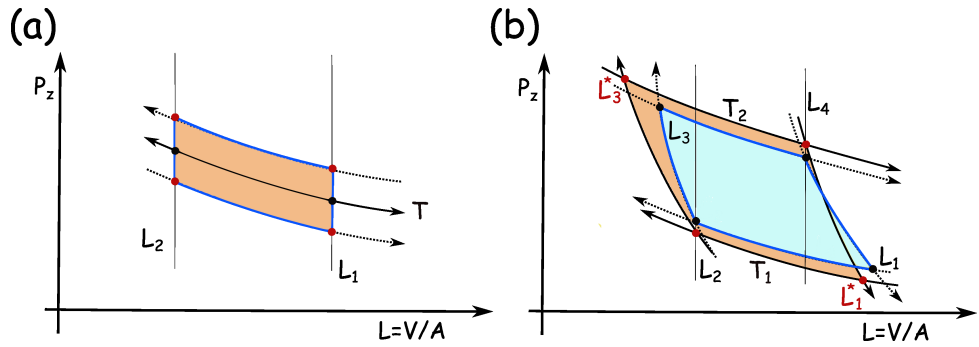}
\caption{
Cartoons of (a) a irreversible isothermal cycle and (b) an irreversible Carnot Process.
With irreversibility in play one can get isothermal cycles, since the pressure $P_z$ the piston
feels during compression is slightly higher than the pressure $P$ against the piston walls, while during expansion
the reverse is true. The deviation of $P_z$ from $P$ increases with the piston speed and vanishes for infinitely slow pistons.
The isothermal cycle for temperature $T$
cannot be run in reverse direction and the orange interior represents the amount of irreversible work required for performing the cycle.
In (b) the adiabatic work contributions cancel each other,
since as in the reversible case, the work corresponds to the difference of the internal energy of the gas.
However, for finite cycle periods
$0<\tau<\infty$ the irreversible character of the adiabatic parts shows in the extremal
positions $L_1$ and $L_3$ of the cycle, if we keep the positions $L_2$ and $L_4$ fixed, where the process switches
from an isothermal to an adiabatic part of the cycle. When pushing the piston in, the internal energy increases faster so that $L_3<L^*_3$, where $L^*_3$ is the minimal cycle position of the reversible process. When the piston pushes out,
it takes the adiabatic process longer to reach $T_1$ and therefore $L_1>L^*_1$
The isothermal irreversibility shows in $P_z$, the pressure on the piston,
which is reduced with respect to $P$, the pressure on the cylinder walls, when pulling the piston out, while $P_z>P$ when pushing the piston in.
As a consequence, the amount of irreversible work lost in one cycle, $\Delta W^{\rm irrev}_{\rm Carnot}$, is represented by the orange area, while
the work gained in one irreversible cycle is depicted in blue.
%
}
\label{fig3}
\end{figure}

\vspace*{-5pt}

\subsection{The adiabatic dilation}

Let us assume that we compress (expand) adiabatically, moving the piston from position $L_1$ to some position $L_2$ in a timespan $\tau$. If at position $L_1$ the working gas has temperature $T_1$, we may ask the question
for which piston position $L_2<L_1$
($L_2>L_1$)
the working gas will reach a given temperature $T_2>T_1$ ($T_2<T_1$).
It is not difficult to write down the exact equations for the problem, in fact it is Eq. (\ref{adiabatic2}), but one cannot in general
solve this equation explicitly as it is of higher order polynomial rank.
Fortunately we can solve it in a linear approximation.
By assuming that cycle periods are large enough for the piston position $L_2=L^*_2+\Delta L_2$, to vary only by a comparably
small, $\Delta L_2\ll |L_2-L_1|$, where $L^*_2$ is the would be piston position if the system would undergo a reversible adiabatic process
ending in temperature $T_2$,
one can compute $\Delta L_2$ in a perturbative manner.
This means that for $L^*_2$ the typical equation for reversible adiabatic
curves,
\begin{equation}
\frac{\bar p_z(T_2)}{\bar p_z(T_1)} \ = \ \sqrt{\frac{T_2}{T_1}} \ = \ \left(\frac{L_1}{L^*_2}\right)^\alpha \,,
\label{revadiabat}
\end{equation}
must hold.
Therefore, substituting $L^*_2+\Delta L_2$ for $L_2$ in Eq.~(\ref{adiabatic2}) and expanding
the equation to first order in $\Delta L_2$ leads to the linearised equation
\begin{equation}
\Delta L_2 \ = \ -\left[  \frac{4\alpha\tau}{m L^*_2} \frac{ \bar p_z(T_1)}{D_z^*(L^*_2|L_1)}\ -\ \frac{\partial}{\partial L_2} \log D_z^*(L_2|L_1)|_{L_2=L^*_2}
\right]^{-1}.
\label{firstorder1}
\end{equation}
By repeated usage of Eq.~(\ref{revadiabat}) we can transform Eq.~(\ref{firstorder1}) into
\begin{equation}
\frac{\Delta L_2}{L_1} \ = \ \left(\frac{T_1}{T_2} \right)^{\frac{1}{2\alpha}}
\left[ \frac{4\alpha\tau}{m L_1}B(T_1|T_2)    -A(T_1|T_2)  \right]^{-1},
\label{firstorder2}
\end{equation}
with
\begin{equation}
A(T_1|T_2) \ = \ \frac{1}{2}\frac{ 1 }{1-\left({T_2}/{T_1} \right)^{\frac{1}{4\alpha}}}
\ + \ \frac{\alpha+\frac12}{1-\left({T_2}/{T_1} \right)^{\frac{1}{4\alpha}+\frac12}},
\label{firstorder2A}
\end{equation}
and
\begin{equation}
B(T_1|T_2)\ = \ \frac{\bar p_z(T_1)T_2^{\frac{1}{2\alpha}}}{D_z^*(T_1^{\frac{1}{2\alpha}} | T_2^{\frac{1}{2\alpha}} )  }.
\label{firstorder2B}
\end{equation}

Let us note that $\Delta L_2$ from Eq. (\ref{firstorder2}) is positive for all phenomenologically relevant values of $L$ and $T$.
This can be made plausible by observing that $B(T_1|T_2)>0$ and, for compression, also $A(T_1|T_2)>0$, hence $\Delta L_2>0$.
This is obviously true for an arbitrarily fast moving piston. During expansion $A<0$, and $\tau B+A>0$ only for sufficiently large $\tau$.
This means that $\Delta L_2>0$ remains true also for slow pistons and, as it turns out, for typical phenomenological piston speeds up to
observed idle speed.

Now we have all ingredients needed to tackle Carnot-like irreversible heat engines.

\section{Carnot-like  irreversible heat engines}\label{sec:carnot}

\begin{figure}[h]
   \includegraphics[width=5.5in]{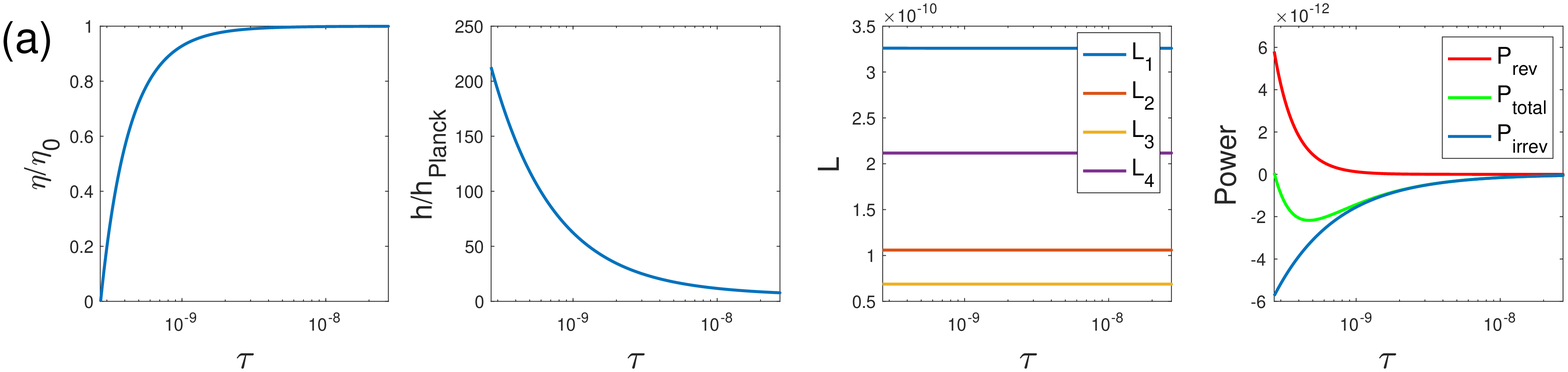}
  \\
   \includegraphics[width=5.5in]{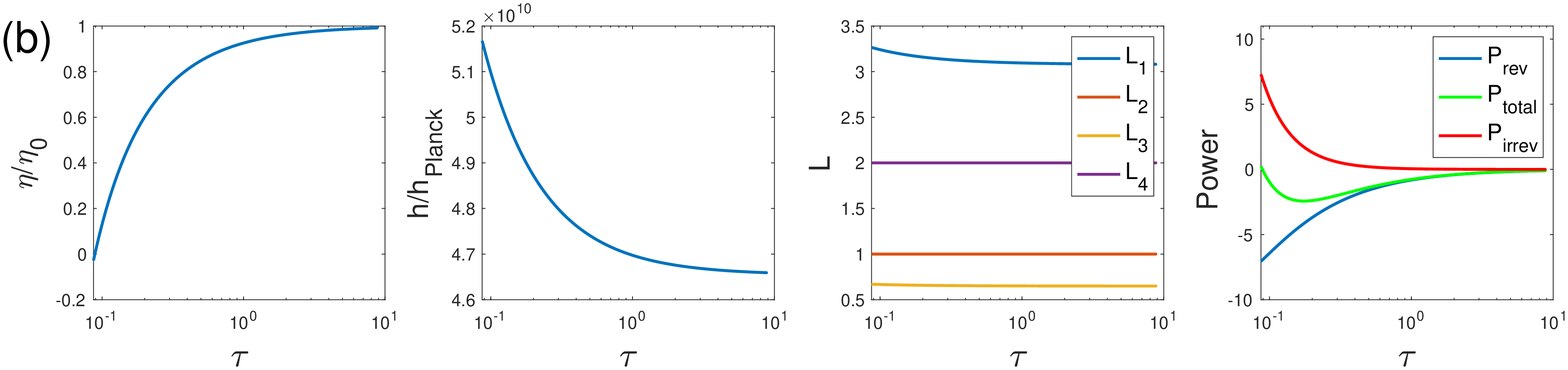}
%

\caption{Basic thermodynamic features of irreversible engines are provided by two examples of heat engines that have mono-atomic $^4$He as a working gas and where we fixed the ratio $L_4/L_2=2$. The engines are driven by two heat baths with respective temperatures $T_1=300$ K and $T_2=400$ K. Engine (a) is microscopic, containing only one $^4$He atom, and  $L_2=14\ 10^{-11}$ m being approximately the Van der Waals radius of $^4$He. It reaches idle speed for approximately $\tau=2.7\ 10^{-10}$ s and its maximum power output at $\tau=4.7\ 10^{-10}$ s.
Engine (b) is macroscopic, with
$L_2=1$ m and a cross-section area of $0.01$ $m^2$. The working gas in the cylinder has normal pressure of $1$ bar with the piston in position $L_4$, which implies that there are about $5.38\ 10^{20}$ $^4$He atoms in  the cylinder.
The engine reaches idle speed for approximately $\tau=0.087$ s and its maximum power output at $\tau=0.175$ s.
}
\label{fig:examples}
\end{figure}

Let us consider a Carnot-like irreversible cycle $L_1\stackrel{\tau_{12}}{\rightarrow}L_2
\stackrel{\tau_{23}}{\rightarrow}L_3\stackrel{\tau_{34}}{\rightarrow}L_4\stackrel{\tau_{41}}{\rightarrow}L_1$, compare
figure (\ref{fig3}), where $W|_1^2(\tau_{12})$ and $W|_3^4(\tau_{34})$ are isothermal work contributions, coupled to heat baths with temperature $T_1$ and $T_2$ respectively ($T_2\geq T_1$). The cycle period is $\tau=\tau_{12}+\tau_{23}+\tau_{34}+\tau_{41}$. The paths $L_2\to L_3$ and $L_4\to L_1$ are the adiabatic parts of the cycle. The first thing we may note about the irreversible Carnot cycle is that,
just as in the reversible case, the adiabatic work contributions cancel  each other.
This is due to $\Delta W = \Delta U$, i.e. the work completely determines the change of the internal energy,
and clearly $\Delta U|_2^3=-\Delta U|_4^1$.
However, the irreversible character of finite-time adiabatic processes has another consequence.
During compression from $L_2\to L_3$ the internal energy increases faster than in the reversible case,
where compression starts at $L_2$ and ends at $L_3=L^*_3+\Delta L_3$, where $L^*_3$ is the would-be equilibrium
position of the piston. $L_3>L^*_3$ is the piston position where the working gas reaches the internal energy that corresponds to the temperature $T_2$ of the hot reservoir, which is the point where the process has to switch from adiabatic to isothermal. Similarly, in the case of irreversible adiabatic expansion  from $L_4\to L_1$ the internal energy decreases less steeply, again because of irreversible work contributions, and the expansion starting at $L_4$ ends up at $L_1=L^*_1+\Delta L_1$, where $L_1>L^*_1$ is the piston position where the irreversible process reaches $T_1$, the temperature of the cold heat bath. $L^*_1$ is again the respective would-be equilibrium piston position.

We can now use Eq.~(\ref{firstorder2}) to compute the adiabatic dilation of the extremal points of the cycle describing the path of the irreversible Carnot-like heat engine and Eq.~(\ref{isowork2a}) to compute the total work
as the sum of the two respective isothermal work contributions. It has to be noted that each path segment can be associated with individual times $\tau_{12}$, $\tau_{23}$, $\tau_{34}$, $\tau_{41}$, which together yield the cycle period
\begin{equation}
\tau \ = \ \tau_{12} \ + \ \tau_{23} \ + \ \tau_{34} \ + \ \tau_{41}\, .
\label{tauconstraint}
\end{equation}

In a last step one can determine the four times $\tau_{ij}$ in such a way, that for some fixed $\tau$, the irreversible work losses $\Delta W^{\rm irrev}_{\rm Carnot}$
(see the orange area in Fig. (\ref{fig3})) become minimal. This can be achieved by varying the
total work created in the irreversible cycle with respects to the times $\tau_{ij}$, conditioned
to a fixed value of $\tau$ using the method of Lagrangian multipliers. This yields a set of fixed
point equations that can be numerically solved in an iterative way, starting from the uniform
initial condition  $\tau_{ij}=\tau/4$.
See, for instance, Fig.~(\ref{fig:examples}) for examples of a microscopic and a
macroscopic heat engine.
As our computations above suggest, and those examples confirm, the minimal possible value of
$\tau\, \Delta w^{\rm irrev}_{\rm Carnot}$ over all paths consistent with the defining process parameters ($L_2$, $L_4$, $T_1$, $T_2$, and $\tau$) is a positive constant $\hat h_{\rm Carnot}$ (see below)  that
is attained in the limit of large cycle-times, $\tau\to\infty$, when the efficiency of the irreversible heat engine
\begin{equation}
\eta \ = \ 1+\frac{\Delta w|_1^2}{\Delta w|_3^4}\,,
\label{carnotefficiency}
\end{equation}
approaches the value of the reversible Carnot process, i.e. $\eta_0=1-T_1/T_2$, implying $\eta/\eta_0 = 1$.
In passing we should stress that the formula Eq.~(\ref{carnotefficiency}) is in its spirit very different from the so-called
internal efficiency formulas often used in engineering applications~\cite{Pulkrabek}.
Particularly, finite heat transfer plays no role in our considerations.

As a consequence, we get that for 
any of the consistent paths $\zeta\, \equiv\, t\to L(t)$ the uncertainty relation
\begin{equation}
\tau\, \Delta w^{\rm irrev}(\zeta)\ \geq\ {\rm minval}(\tau\, \Delta w^{\rm irrev}_{\rm Carnot})\ \equiv\ \hat h_{\rm Carnot}\ \geq\ 0\, .
\label{uncertaintyX}
\end{equation}
It is not difficult to recognize that this uncertainty relation can be extended to general irreversible thermodynamic heat engines by imagining, similarly as in reversible thermodynamics, that we tessellate the irreversible cycle by infinitesimal Carnot cycles, which at the cycle boundary traverse the respective isothermal and adiabatic path elements with cycle's local traversing speed.
Moreover, from Fig~\ref{fig:examples} we see that the minimal value $\hat h_{\rm Carnot}$  arises in the quasi-static limit, i.e. the limit $\tau\to \infty$. This limit can be computed explicitly, and
after some simple algebra
one obtains the expression for  $\hat h_{\rm Carnot}(L_2,T_1;L_4,T_2)$ in the form
\begin{equation}
\hat h_{\rm Carnot}=(\vartheta+\nu)(\vartheta+\xi)\, ,
\end{equation}
where the functions $\vartheta$, $\nu$, and $\phi$ are given by
\begin{equation}
\begin{array}{lcl}
\vartheta &=& \frac{1}{2} D_z(  L_2 T_1^{\frac{1}{2\alpha}} , L_4 T_2^{\frac{1}{2\alpha}}  )^{\frac12}   \left[ \left( \frac{\bar p_z(T_1)}{T_1^{\frac{1}{2\alpha}}}   \right)^{\frac{1}{2}}+\left( \frac{\bar p_z(T_2)}{T_2^{\frac{1}{2\alpha}}} \right)^{\frac{1}{2}}\right]\! ,\\
\nu &=& \frac12D_z^*(T_1^{\frac{1}{2\alpha}} , T_2^{\frac{1}{2\alpha}})^{\frac12}\left[ \bar p_z(T_2)\left( \frac{L_2}{\bar p_z(T_1)T_2^{\frac{\alpha+1}{2\alpha}}} \right)^{\frac{1}{2}} + \bar p_z(T_1)\left( \frac{L_4}{\bar p_z(T_2)T_1^{\frac{\alpha+1}{2\alpha}}}\right)^{\frac{1}{2}}\right]\! ,\\
\xi &=& \frac{m}{2\alpha}D_z^*(T_1^{\frac{1}{2\alpha}} , T_2^{\frac{1}{2\alpha}})^{\frac12}\left[ \left( \frac{L_2}{\bar p_z(T_1)T_2^{\frac{\alpha+1}{2\alpha}}}\right)^{\frac{1}{2}} + \left(\frac{L_4}{\bar p_z(T_2)T_1^{\frac{\alpha+1}{2\alpha}}}\right)^{\frac{1}{2}}\right]\! .
\end{array}
\end{equation}
If one keeps the ratio $L_2/L_4$ fixed and  uses, for instance, $L=L_2$ as the characteristic length scale  of the cycle, one may notice that $\hat h_{\rm Carnot}\propto L$, i.e. the amount of irreversible work per cycle and per a single degree of freedom scales linearly with the size of the engine.

To put some flesh on the bare bones,  we consider two examples of heat engines  running upon
a working substances modeled by a mono-atomic ideal gas, such as  $^4$He. In particular, we inspect: (a) a microscopic engine operating at the typical length scale of $^4$He atom, and (b) a macroscopic engine operating at the scale $\sim 1~m$.
In both these cases we fix the ratio $L_4/L_2 =2 $, i.e. roughly speaking, we increase localization of a gas particle by one bit.
The engines are driven by two heat baths with respective temperatures $T_1=300$ $K$ and $T_2=400$ $K$. Engine (a) is microscopic, containing only one $^4$He atom, and
$L_2= 2.8$~{\AA},
which is approximately
the Van der Waals diameter of $^4$He.
The engine reaches the idle speed at approximately $\tau=2.7\ 10^{-10}~s$ and its maximum power output at $\tau=4.7\ 10^{-10}~s$.
Moreover, $\hat h_{\rm Carnot}=4.93\ h_{\rm Planck}$ (with $ h_{\rm Planck}=6.626\ 10^{-34}$ $Js$), i.e. the irreversible mechanical energy loss for the microscopic localization of particles is of the same order of magnitude as quantum phenomena.
As a side-note, those characteristic cycle frequencies lie well in the micro-wave band which we use on a daily basis in our kitchens to transfer energy into organic matter at the molecular level.
For the macroscopic engine (b) we choose $L_2=1$ $m$ and a cross-section area of $0.01$ $m^2$. The amount of the ideal gas in the cylinder is chosen so that we have normal pressure of $1$ bar with the piston being in position $L_4$, which implies that there are about $5.38\ 10^{20}$ $^4$He atoms in  the cylinder. The engine reaches idle speed at approximately $\tau=0.087$ $s$ and its maximum power output at $\tau=0.175$ $s$. Moreover, $\hat h_{\rm Carnot}=4.66\ 10^{10}\ h_{\rm Planck}$, i.e. the effect of mechanical irreversibility also is macroscopic.
The predictions based on the irreversible properties of mechanical energy transfer at the piston are depicted on Fig.~\ref{fig:examples}.

As a simple consistency check of the validity of our initial assumption that the piston velocity is much smaller than
the characteristic particle velocity we compute the ratio $|\dot L|/ \bar v_z$ at the limiting idle speed, i.e. the speed where the efficiency of the irreversible cycle vanishes.  Explicit computations give the value $9.97\times 10^{-04}$ in the microscopic case (a) and $0.029$ in the macroscopic case (b).
These results imply that corrections to the first order approximation considered in this paper would typically
be of the same magnitude, meaning that we may expect an error of the magnitude
$\hat h_{\rm Carnot}/\ h_{\rm Planck}\sim 4.93\pm 0.005$ for (a) and
$\hat h_{\rm Carnot}/\ h_{\rm Planck}\sim (4.66 \pm 0.15)\ 10^{10}$ for (b).


\section{Conclusion}

In this paper we have analysed how \textit{irreversible work} contributions to real heat engines are created at the
\textit{mechanical} interface via elastic collisions of ideal-gas particles with a finite-mass piston controlling
the volume of the gas.
It turns out that the force required to move the piston at a non-zero speed agrees with an acceleration of the piston between particle collisions that breaks the isotropy of particle velocities heading towards the piston with respect to those rebounding from the piston.
Macroscopically this corresponds to a slight difference in the gas pressure that the moving piston experiences (sometimes called instantaneous pressure) relative to the static situation (internal pressure).
This, in turn, is responsible for the mechanical work that is irreversibly lost in a finite-time-cycle process; either to the heat bath (for isothermal process) or to an increased internal energy (for adiabatic process).
In the limit of infinitely slow piston speed, the amount of irreversible energy required vanishes and one recovers the predictions of quasi-static theory.
%
However, an important property, the \textit{dissipative action} , survives
the limit  and  $\tau\Delta w_{\rm irrev}>\hat h_{\rm process}>0$  for all $\tau\geq 0$, i.e. the limit of the
product of the cycle period $\tau$ and the amount of irreversible work $\Delta w_{\rm irrev}$ per degree of freedom generated in one cycle, cannot vanish in the quasi-static limit since it is bounded from below by the constant $\hat h_{\rm process}$. In particular, this implies that in the quasi-static limit
$\Delta w_{\rm irrev} = {\cal{O}}(1/\tau)$.
The aforementioned uncertainty relation resembles the celebrated Heisenberg (or better Tamm--Mandelstam) time-energy uncertainty relation.
Despite the different operational meanings, the presented uncertainty relation can be viewed as a classical analogue of corresponding quantum-mechanical relations for periodic systems, namely
that a system's period (e.g. neutrino oscillation period)
provides a fundamental bound on energy degradation~\cite{BJS}. Moreover, for irreversible cycles at atomic scales also the process specific constants $\hat h_{\rm process}$ are of the same order of magnitude as $h_{\rm Planck}$, implying a comparability of irreversible thermodynamic processes and quantum effects at this scale.
The uncertainty relations provide also an interesting connection with information theory. To this end we
consider a single-particle ``gas'' confined within a vessel. To increase the localization of the single particle
from a volume $V$ to a volume $V/2$ corresponds to gaining one more bit of information
on the position of the particle~\cite{Feynman,Szilard}. Erasing this bit means expanding back from $V/2$ to $V$. As a consequence, writing and erasing a bit of information within a time $\tau$ comes at an irreversible energy cost per degree of freedom that is bounded from below by $\hat h_{\rm process}/\tau >0$ that depends on the physical scale of the process and its characteristic temperatures. Or in other words,
if an observer loses information about a physical system, the observer loses the ability to extract work from that system. In this sense this might be viewed as a generalization of Landauer's principle~\cite{Landauer} to two heat baths.

At macroscopic scales our time-energy uncertainty principle implies that thermodynamic engines inevitably possess an idle speed, meaning that they run with a characteristic speed without exterior workload. At this point the efficiency of the engine is zero and all work produced is irreversibly spent on the process  running through the cycle in a finite time. Moreover, while the most efficient process is unavoidably the reversible quasi-static process, the maximum power output is reached for another characteristic cycle period, where the relation between number of cycles performed per time unit and the ensuing reduction of work efficiency (due to irreversible energy production) are optimal.

Our  analysis rests on the ``dullest'' thinkable situation;
(a) an ideal gas as working medium that essentially remains
homogeneous and isotropic in the cylinder, (b) an idealized piston gliding frictionless in this
cylinder,
and (c) instantaneous heat transfer between heat bath and working gas.
This means that we disregard a number of phenomena that in general make real processes far richer and more ``interesting'' with contributions adding to irreversible work requirements; pressure waves and
resonance phenomena, spectrum of phonons or
type of vibrational modes in the piston wall, coupling to the heat-bath, etc.
However, all those processes can only add to irreversible work and increase the effective value of
the process specific action constant,
but they do not cancel the effect of momentum transfer of molecules with a moving piston that we
have mathematically
analysed
in this paper.

We should also point out that in principle it is possible to disentangle various irreversible contributions and determine the relative magnitude and characteristics of the discussed (mechanistic)
effect by adapting the outlined theory to pertinent  experimental data;
for instance, data obtained by versions of the R{\"u}chardt experiment~\cite{ruechardt,mungan},
or other experiments that can supply information on the difference between instant and internal gas pressure.
However, due to the path dependence of the force equation~(\ref{workfunctionalA}), adapting
the presented theory to the geometry and piston dynamics of particular experimental setups becomes
a non-trivial task that certainly goes beyond the scope of this paper.
The details  to such ends  will be discussed elsewhere.

The presented analysis of the mechanical interface in our ideal gas framework prompts a number of interesting questions. Here is a partial list of them. First, we have considered a very simple geometry of the confining vessel.
So, what role plays the shape of the vessel and can the results be formulated in a shape independent manner?
Second, how would the inclusion of finite-heat-transfer coefficients modify our conclusions.
Third, how essential is the ideal gas as a working medium.
What about photons that are relativistic or Van der Waals gas particles that are not point like?
Fourth, how do the characteristic piston speed, i.e. the piston speed at the idle speed boundary, and
for the speed for maximal power output functionally depend on and compare to typical gas-particle velocities.
Can one go easily beyond the first order approximation employed here?
Fifth, do our results generalize to other non-mechanical interfaces such as electromagnetic or
chemical interfaces.
Finally, one could ask if (or to what extend)  the uncertainty relation (\ref{uncertaintyX}) is a
consequence of entropic inequalities used in stochastic thermodynamics~\cite{SR}
or vice versa.
At this point  we should perhaps
emphasise that the notions of entropy or entropy production
do not enters our analysis. Similarly, as in reversible thermodynamics or in the example of the Curzon--Ahlborn cycle~\cite{CurzonAhlborn,BrownFA}, cyclic thermal processes imply the notion of entropy and not vice versa. In a sense, the reversible cycle is a more primitive concept than thermodynamic (i.e., Clausius) entropy. Could not a similar line of
reasoning hold also on the level of irreversible cycles, namely could not the ensuing time-energy
uncertainty relations imply entropic inequalities?

\vskip6pt

\enlargethispage{20pt}

\ethics{We declare we have no competing interests}

\dataccess{This article has no additional data.}

\aucontribute{Both authors jointly discussed, conceived and wrote the manuscript}

\competing{The authors declare that they have no competing interests.}

\funding{R.H. was  supported by the Austrian Science Fund (FWF) under Project No. I3073 and P.J.  was  supported  by the Czech  Science  Foundation Grant No. 17-33812L.}

\ack{Authors are grateful to Stefan Thurner and Jan Korbel (both from CSH in Wien) for fruitful discussions.
}



\end{document}


\title{Supplemental Material for ``Time-Energy Uncertainty Principle for Irreversible Heat Engines''\\
(DOI: 10.1098/rsta.xxxx.xxxx)}

\author{
Rudolf Hanel$^{1}$ and Petr Jizba$^{2}$}

\address{$^{1}$Section for Science of Complex Systems, Medical University of Vienna, Spitalgasse 23, 1090 Vienna, Austria\\
$^{2}$FNSPE, Czech Technical University in Prague, B\v{r}ehov\'{a} 7, 115 19, Praha, Czech Republic}



\corres{Petr Jizba\\
\email{p.jizba@fjfi.cvut.cz}}
\maketitle

\noindent {\bf Note:} equations and citations that are related to the main text are shown in red.

\subsection*{The average particle velocity reaching the piston}

In order to derive {\cor Eq.~(2.5)} in the main text, one can proceed as follows. First we note that
the ``lower'' piston velocity $\underline u(t)$, directly after a collision at time $t$, with a particle
having a $z$ velocity component, $v_z$, and the ``higher'' piston velocity $\underline u(t)$, directly
before the same collision, are related, if we assume an elastic collision between particle and piston taking place:
\begin{equation}
\underline u(t) \ = \ \frac{2m}{M+m}v_z\ +\ \frac{M-m}{M+m}\overline u(t)\, ,
\label{ellastic_coll}
\end{equation}
where $M$ is the piston mass and $m$ the particle mass.
If we accelerate the piston with a constant acceleration $a(t)$ between the collision event at time $t$ and
the next collision at time $t+\Delta t$, then we also have $\overline u(t+\Delta t)=\underline u(t)+a(t)\Delta t$.
Moreover, by defining the macroscopic piston velocity $\dot L(t)=(\overline u(t+\Delta t)+\underline u(t))/2$
one also gets $\overline u(t+\Delta t)=\dot L(t)+\frac12 a(t)\Delta t$. We now have two equations for
$\underline u(t+\Delta t)$. First, using Eq.~(\ref{ellastic_coll}), we get
\begin{equation}
\underline u(t+\Delta t) \ = \ \frac{2m}{M+m}v_z\ +\ \frac{M-m}{M+m}\overline u(t+\Delta t)\, .
\label{u1}
\end{equation}
Second, we know from above, that also
\begin{equation}
\underline u(t+\Delta t) \ = \ \dot L(t+\Delta t)-\frac{1}{2} a(t+\Delta t) \Delta t'\, ,
\label{u2}
\end{equation}
where $ \Delta t'$ is the time elapsing between second and third collision.
From those two equations, defining the macroscopic piston
acceleration $\ddot L(t)=[\dot L(t+\Delta t)-\dot L(t)]/\Delta t$, it follows that
\begin{equation}
\frac{2m}{M+m}\left[ v_z-\dot L(t) \right] \ = \ \left[\ddot L(t) - \bar a(t)\right] \Delta t \, ,
\label{lemma1}
\end{equation}
where
\begin{equation}
\bar a(t) \ \equiv \ \frac{1}{2}\left[ \frac{M-m}{M+m} a(t)+ a(t+\Delta t) \frac{\Delta t' }{\Delta t}\right] \, .
\label{lemma1x}
\end{equation}
We see that for a particular sequence of events the variable $\bar a(t)$ will be fluctuating. If, however, we control the external instant force on the system smoothly and slowly, we can assume that for a macroscopic
time resolution on average $a(t) = a(t+\Delta t)$ and $\Delta t' = \Delta t $, so that we can replace the microscopic values  by their appropriate macroscopic averages. From this it follows that
\begin{equation}
\bar a(t) \ = \ \frac{M}{M+m} a(t)\, ,
\label{lemma1xx}
\end{equation}
i.e. we get a multiplicative correction of the acceleration which vanishes in the heavy piston limit $M\gg m$.
Now, using that for subsequent collisions it is true that the distance between piston and particle is given by
$\Delta z = (v_z-\dot L(t))\Delta t$. Inserting this into Eq.~(\ref{lemma1}) we obtain
\begin{equation}
\frac{2m}{M+m}\left[ v_z-\dot L(t) \right]^2 \ = \ \left[\ddot L(t) - \bar a(t)\right] \Delta z \, ,
\label{lemma2}
\end{equation}
which is equivalent to {\cor Eq.~(2.5)} in the main text. We point out that this equation is true
for two subsequent collision events, where the second colliding particle with velocity $v_z$  is separated from the piston by a distance $\Delta z$ exactly at the moment the first particle collides. In the next step we have to appropriately replace the values $v_z$ and $\Delta z$ by average values.

If, between $v_z$ and $v_z+dv_z$ one finds $dN(v_z)$ gas particles, then the
average $z$-distance between those particles would be $\Delta z=L/dN(v_z)$ then the average time that passes between such particles colliding with the piston is
$\Delta t=\Delta z/(v_z-\dot L)$
and, as a consequence, the ensuing average frequency density  of collisions with the piston is
\begin{equation}
d\varphi(v_z) \ = \ \frac1L(v_z-\dot L)dN(v_z)\, .
\end{equation}
If the piston moves slowly in comparison with a typical gas-particle speed the particle velocities will
still follow a Maxwell--Boltzmann distribution (except for the particles rebounding from the
piston, which are assumed to have enough time to equilibrate with the bulk of the gas before
they return to the piston) and hence
\begin{equation}
\frac{dN(v_z)}{N} \ = \ \left(\frac{\beta m}{2\pi}\right)^{1/2}e^{-\frac{\beta m}{2}v_z^2}dv_z\,.
\end{equation}
The average total particle collision rate (i.e., average frequency of collisions) with the piston is therefore given by
\begin{equation}
\varphi \ = \ \int_{\dot L}^\infty d\varphi(v_z)=\frac{N}{2L}\left(\sqrt{\frac{2}{\pi}}\bar v_z-\dot L\right)+R_1\, ,
\label{meanrateSI}
\end{equation}
where one may note that, $R_1=\int_0^{\dot L} d\varphi(v_z)$, can be ignored as long
as one only wants to obtain an approximation in the first-order  in $\dot L/\bar v_z$.
The average particle velocity that is sampled by the piston is given by
\begin{equation}
\langle v_z \rangle_{\rm piston} \ = \ \frac1\varphi \int_{\dot L}^\infty d\varphi(v_z) v_z
\ = \ \frac{\bar v_z-\sqrt{\frac{2}{\pi}}\dot L}{\sqrt{\frac{2}{\pi}}\bar v_z-\dot L}\bar v_z + R_2\,,
\label{meanvzSI}
\end{equation}
where $R_2=\int_0^{\dot L} d\varphi(v_z) v_z$ again can be ignored in the first-order approximation.

It follows that the value of $\overline{\Delta t}$,  i.e., the average time elapsing between two collisions, is given by $\overline{\Delta t}=1/\varphi$.
It now becomes possible to make the Ansatz $\Delta z=(v_z-\dot L)/\varphi$ for
the colliding particle-piston distance in reference to the average inter collision time $\overline{\Delta t}$.
Inserting this result into Eq. (\ref{lemma2}) gives us
\begin{equation}
\frac{2m}{M+m}\left[ v_z-\dot L(t) \right] \ = \ \frac{1}{\varphi}\left[\ddot L(t) - \bar a(t)\right] \, ,
\label{lemma3}
\end{equation}
Integrating both sides with respect to the measure $d\varphi(v_z)$ in the respective $v_z$ range, $\dot L$ to $\infty$
and using Eq.~(\ref{lemma1x}) further implies
\begin{equation}
\frac{m}{M}\frac{N}{L}\bar v_z^2\left[ 1 - 2\sqrt{\frac{2}{\pi}}\frac{\dot L}{\bar v_z}
+ \left(\frac{\dot L}{\bar v_z} \right)^2 \right] \ = \ \left(1+\frac{m}{M}\right)\ddot L(t) - a(t)  \, ,
\label{lemma4}
\end{equation}

Finally, we need to disentangle the work contribution $dW_{\rm gas}$ we obtain for compressing or relaxing the gas from the work contributions we obtain from accelerating the piston itself. We note that the total work
necessary for moving both piston and compressing/relaxing gas is given by
\begin{equation}
dW_{\rm gas\ \&\ piston}(t) \ = \ Ma(t)dL(t)  \, .
\label{gaspiston}
\end{equation}
As a consequence, the amount of work related to gas compression or expansion is given by $dW_{\rm gas}=dW_{\rm gas\ \&\ piston}(t)-M\ddot L(t)dL(t)$, i.e.
\begin{equation}
dW_{\rm gas} \ = \ -mN \frac{dL}{L}\bar v_z^2\left[1-2\sqrt{\frac{2}{\pi}}\frac{\dot L}{\bar v_z} + \left(\frac{\dot L}{\bar v_z}\right)^2 +\frac{\ddot L L}{N\bar v_z^2} \right] \, .
\label{gas}
\end{equation}
If we want to approximate the equation to the first order in $\dot L/\bar v_z$, which is justified if $1\gg |\dot L|/\bar v_z$,
then we still need to know whether we can disregard the remaining term stemming from piston acceleration,
\begin{equation}
\frac{|\dot L|}{\bar v_z} \ \gg \ \left| \frac{\ddot L L}{N\bar v_z^2} \right|\, .
\label{condition}
\end{equation}
If we assume this to be true, then we learn in the main body of the paper that the optimal path between extremal
piston positions, $L_1$ and $L_2$, is given by paths with constant acceleration
\begin{equation}
\ddot L=2\left(\frac{\sqrt{L_1}-\sqrt{L_2}}{\tau}\right)^2\, .
\label{acceleration}
\end{equation}
If we insert this expression for $\ddot L$ into Eq. (\ref{condition}), then we obtain, using
$\tau|\dot L|\sim 2(L_{\max}-L_{\min})$, that
$N\tau\bar v_z \ \sim \ 2(L_{\max}-L_{\min})N\bar v_z \ \gg \ |\dot L|(\sqrt{L_{\max}}-\sqrt{L_{\min}})\sqrt{L_{\max}}$, i.e.
\begin{equation}
N\left(1+\sqrt{\frac{L_{\min}}{L_{\max}}}\right) \ \geq \ 1 \ \gg \ \frac{|\dot L|}{\bar v_z}\, ,
\end{equation}
which is fulfilled, since we made the assumption of slow piston velocities ($1 \ \gg \ |\dot L|/\bar v_z$), and
justifies the first order equation for $dW_{\rm gas}$,
\begin{equation}
dW_{\rm gas} \ = \ -mN \frac{dL}{L}\bar v_z^2\left(1-2\sqrt{\frac{2}{\pi}}\frac{\dot L}{\bar v_z} \right) \, ,
\label{firstordergas}
\end{equation}
which we use throughout the main body of the paper, for the slow piston  operating range.